\documentclass[12pt]{article}
\usepackage{appendix}
\usepackage{epsfig}
\usepackage{amsmath}
\usepackage{amssymb}
\usepackage{bm}
\usepackage{hyperref}
\usepackage[normalem]{ulem}
\usepackage{color}
\usepackage{cite}

\newcommand{\be}{\begin{equation}}
\newcommand{\ee}{\end{equation}}
\newcommand{\ba}{\begin{eqnarray}}
\newcommand{\ea}{\end{eqnarray}}

\newcommand{\hrho} {\hat{\rho}}
\newcommand{\hq}{\hat{q}}
\newcommand{\BR}{\mathcal{B}}
\newcommand{\cL}{\mathcal{L}}

\newcommand{\cmu}{c_{1\mu}}
\newcommand{\ch}{c_{1h}}
\newcommand{\chZ}{c_H}

\begin{document}

\begin{center}
\vspace{0.5cm}
    {\Large\bf The $h\to 4 \ell$ spectrum at low $m_{34}$:  \\[0.2cm]
    Standard Model vs.~light New Physics.} \\[0.8cm]
   {\bf Mart\'in Gonz\'alez-Alonso and Gino Isidori}    \\[0.3cm]
   {\em INFN, Laboratori Nazionali di Frascati, I-00044 Frascati, Italy}  \\[1.0cm]
\end{center}

\centerline{\large\bf Abstract}
\begin{quote}
\indent
We analyze $h\to 4\ell$ decays in the kinematical region where the
dilepton pair of low invariant mass ($m_{34}$) is not far from QCD resonances 
($\Psi$ and $\Upsilon$ states). On the one hand, we present precise predictions of the
spectrum within the Standard Model, taking into account non-perturbative QCD 
effects. On the other hand, we illustrate the sensitivity of this spectrum to New Physics models predicting the existence of new exotic light states.  
In particular, we show that parameter regions of models 
relevant to explain the $(g-2)_\mu$ anomaly could be probed in the future by means of 
$h\to 4\ell$ decays.
\end{quote}
\vspace{5mm}

\section{Introduction}

The discovery of the Higgs boson at the LHC~\cite{Higgs} has opened an interesting new chapter of 
phenomenological studies:  the precise investigation of the properties of this new particle,
which appears to be the unique 
massive excitation of a fundamental scalar field. 
So far, the measured couplings of   the Higgs boson ($h$) to Standard Model (SM)  fields are compatible with those expected 
within the SM (see e.g.~Ref~\cite{Giardino:2013bma} and references therein). 
However, the effective interactions of the $h$ particle are still poorly known.


The Higgs field is the only SM field that could have relevant or marginal 
interactions with exotic new states that are singlet under the SM gauge group~\cite{Patt:2006fw}. 
For this reason, it is quite natural to conceive New Physics (NP) models with sizable 
modifications of the $h$ couplings to SM or exotic states, and 
negligible impact in the electroweak precision tests  (in agreement with present data). 
The sensitivity of $h$ decays to physics beyond the SM is further 
strengthened by the measured value of the the Higgs boson mass.  Since $m_h < 2 m_W$,
the potentially leading SM decay modes (to $WW$, $ZZ$, and $t\bar t$) are kinematically forbidden. 
This fact implies an enhanced sensitivity to sub-leading $h$ decay channels ruled 
by small effective couplings. These  include rare SM decay modes (such as 
the semi-hadronic exclusive modes~\cite{Isidori:2013cla,Bodwin:2013gca}), but may 
also include channels that are completely forbidden within the SM (such as lepton-flavor violating modes~\cite{hLFV}
or decays involving new exotic light particles~\cite{Curtin:2013fra}). Both rare and 
forbidden $h$ decay modes could provide an interesting window on 
physics beyond the SM.

In this paper we analyze the possibility to discover rare exotic $h$ decay modes 
in the final states with two pairs of opposite sign light leptons, generically denoted by $h\to 4\ell$. 
In particular, we analyze $h\to 4\ell$ decays in the kinematical region where the
dilepton pair of low invariant mass ($m_{34}$) is not far from QCD resonances 
($\Psi$ and $\Upsilon$ states).  The purpose of the paper is twofold. 
On the one hand, we present precise predictions of the $m_{34}$
spectrum within the SM, taking into account non-perturbative QCD 
effects associated to the quarkonium thresholds. On the other hand, we illustrate the sensitivity of this spectrum 
to NP models, pointing out the natural connection between anomalies in the 
$h\to 4\mu$ channel and NP contributions to $(g-2)_\mu$. 
If the current $(g-2)_\mu$ anomaly is due to the one-loop 
exchange of exotic light mediators (with mass well below $m_h$), 
this easily imply visible deviations from the SM in the 
$m_{34}$ spectrum of $h\to 4\mu$.

While this work was in progress, an extensive discussion about exotic Higgs 
decays has been presented in Ref.~\cite{Curtin:2013fra}. Given the two goals 
outlined above, our analysis is largely complementary with respect to Ref.~\cite{Curtin:2013fra}. The paper is organized as follows: 
in Sect.~\ref{sect:SM} we analyze the $m_{34}$ spectrum within the SM,
with the inclusion of quarkonium effects. Sect.~\ref{sect:NP} is devoted to
explore the connections between $(g-2)_\mu$ and 
$h\to 4\ell$ decays in a few representative NP models.
The results are summarized in the Conclusions. 

\section{The $m_{34}$ spectrum within the Standard Model}
\label{sect:SM}

Within the SM the Higgs decay into two pairs of opposite sign  leptons 
is dominated by the tree-level  amplitude  $h\to Z Z^* \to Z \ell^+\ell^-$, with 
the (quasi) on-shell $Z$ decaying into a $\ell^+\ell^-$ pair ($\ell=e,\mu$). Following the notation introduced in the ATLAS~\cite{Aad:2013wqa}
analyses of these modes, we denote by $m_{12}$ the dilepton invariant mass close  to $m_Z$  and 
by $m_{34}$ the low dilepton invariant mass far from the $Z$ pole ($m^2_{34} \ll m^2_Z$).\footnote{In the notation 
of CMS~\cite{Chatrchyan:2013mxa}, $m_{12} \to m_{Z_1}$ and $m_{34} \to m_{Z_2}$.}
The tree-level decay rate for the (idealized) $h \to Z \ell^+\ell^-$ process  is 
\be
\label{eq:gamma0}
\frac{d\Gamma_0^{\rm SM} (h\to Z\ell^+\ell^-)}{d m_{34}^2} 
= \frac{m^6_Z }{8\pi^3 v^4 m_h } \left[ (g_R^\ell)^2 + (g_L^\ell)^2 \right] ~ \frac{ \lambda(\hq^2,\hrho)}{(m_{34}^2-m_Z^2)^2} \left[ m_{34}^2 + \frac{m_h^4}{12 m_Z^2}~  \lambda^2(\hq^2,\hrho)  \right],
\ee
where $\hrho = m_Z^2/m_h^2$, $\hq^2 = m_{34}^2/m_h^2$, $\lambda(\hq^2,\hrho) = \sqrt{(1 + \hq^2 - \hrho)^2 - 4 \hq^2}$, 
$g_L^\ell = T_3^\ell - Q_\ell s_W^2 $,  $g_R^\ell = - Q_\ell s_W^2$,\footnote{In the following we also use the notation $g_V^f = g_L^f+g_R^f = T_3^f - 2Q_f s_W^2$ and $g_A^f =  g_R^f - g_L^f = - T_3^f$, for both quarks ($f=q$) and leptons ($f=\ell$). Notice that the definition 
of $g_{L,R}^f$ in Ref.~\cite{Isidori:2013cga} is different by a factor of~2. }  and $v=(\sqrt{2} G_F)^{-1/2}\approx$~246~GeV.
The tree-level decay rate for the physical $h\to 4\ell$ decay is described by the convolution of Eq.~(\ref{eq:gamma0}),  
where $\hrho$ is replaced by $m^2_{12}/m_h^2$, with a Bright-Wigner distribution for $m_{12}$ around the $Z$ peak. 
   
\begin{figure}[t]
\centering
\includegraphics[width=0.6\textwidth]{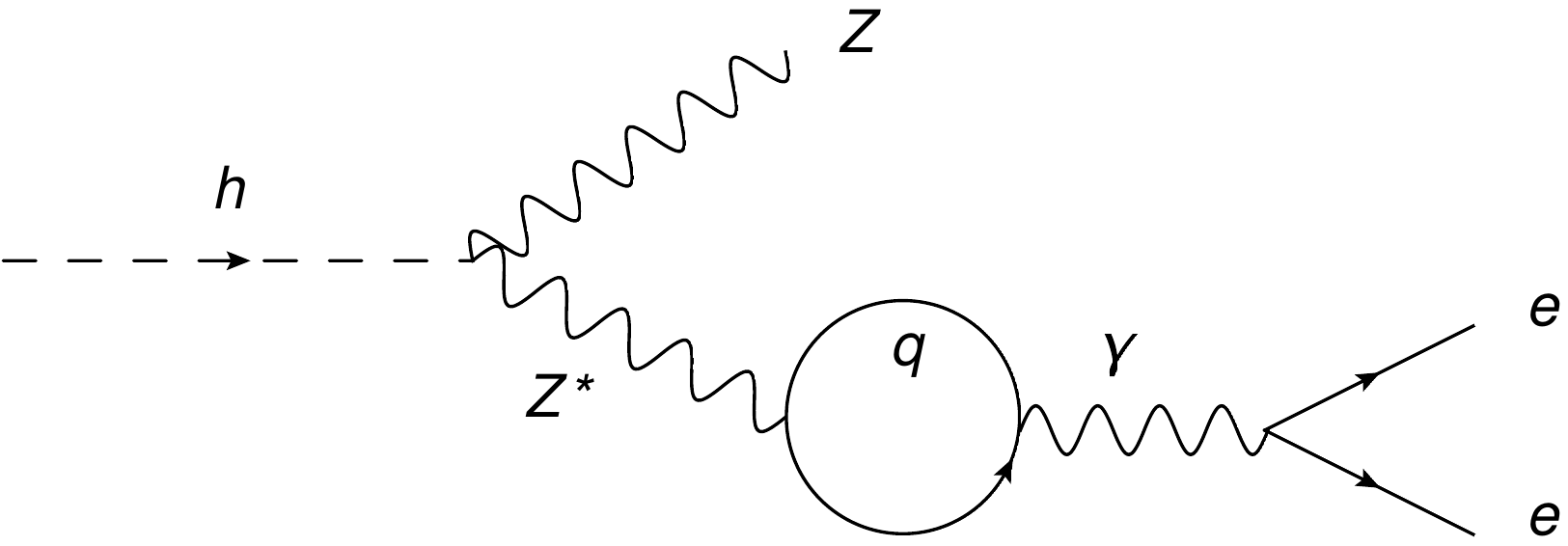}
\caption{One-loop quark contribution to $Z-\gamma$ mixing.} 
\label{fig:one-loop-graph}
\end{figure}

The tree-level expression in Eq.~(\ref{eq:gamma0}) is modified by next-to-leading order (NLO) electroweak corrections.
In general these corrections are tiny, around the few per-mil level (slightly larger in the case of genuine QED effects),
and leads to smooth modification of the $q^2\equiv m_{34}^2$ spectrum.
However, there are two classes of NLO effects generating larger local modifications of the spectrum,
being associated to physical poles in the $q^2$ distribution within (or very close to) the allowed kinematical range:
\begin{enumerate}
\item[I.] the (one-loop) $h\to Z\gamma^*$ effective vertex, which leads to the appearance of a pole at $q^2\to0$;
\item[II.] the distortions of the spectrum due to narrow hadronic resonances contributing to $Z-\gamma$ mixing
(see Fig.~\ref{fig:one-loop-graph}), leading to narrow poles for $q^2\to m^2_{\rm res}$.
\end{enumerate}

The additional contribution of the $h\to Z \gamma ^* \to Z \ell^+\ell^-$
amplitude  leads to the following correction term,
\ba
\frac{d \Gamma_1^{\rm SM} (h\to Z\ell^+\ell^-)}{dq^2} 
&=& \frac{m^6_Z }{8\pi^3 v^4 m_h }  \lambda(\hq^2,\hrho) \left\{ - \frac{  \alpha A_{Z\gamma}^{\rm SM} }{4\pi}  ~
\frac{ Q_\ell (g_L^\ell +g_R^\ell) }{ q^2-m_Z^2} ~ \frac{ m_h^2 (1 -\hat q^2 - \rho)}{m_Z^2}  \right. \nonumber\\
 && \left. + \left( \frac{ \alpha A_{Z\gamma}^{\rm SM} }{4\pi} \right)^2 
\frac{ Q^2_\ell  }{ q^2 } ~ \frac{ m_h^4 [ 3 (1 -\hat q^2 - \rho)^2 -\lambda(\hq^2,\hrho)^2]  }{6 m_Z^4}   \right\}~,
\label{eq:gamma1}
\ea
which cannot be neglected at low $q^2$. Here $A_{Z\gamma}^{\rm SM} = c_W^2 A_W +  (2/3)(3-8 s_W^2) A_t \approx -4.8 $ is the 
reduced one-loop effective vertex dominated by $W$-boson ($A_W\approx -6.5$) and top-quark loops ($A_t \approx 0.3$)~\cite{Cahn:1978nz} (see also Ref.~\cite{Grinstein:2013vsa}). Note that we include both the interference 
term and the modulo square of the $h\to Z \gamma ^* \to Z \ell^+\ell^-$ amplitude. The latter is formally a higher-order correction; 
however, it is the term that leads to the largest local modification in the $q^2$ distribution because of the pole 
at $q^2\to0$. As we discuss in the following, a similar phenomenon happens for  the 
 distortions of the spectrum due to narrow hadronic resonances contributing to $Z-\gamma$ mixing.
  
\subsection{Hadronic contributions to $Z-\gamma$ mixing}

In order to take into account the effect of narrow hadronic resonances in $Z-\gamma$ mixing, 
we introduce the two-point correlation-function $\Pi^{\mu\nu}_{Z\gamma}(q)$, defined as 
\ba
\Pi_{\mu\nu}^{Z\gamma}(q) \equiv  i \int d^4x e^{iqx} \langle 0 | T{ J_\mu^Z(x) J_\nu^\gamma (0)} |0\rangle = - \left( g^{\mu\nu} q^2 - q^\mu q^\nu \right) \Pi_{Z\gamma}(q^2)~,
\label{eq:defcorrelator}
\ea
where $J_\nu^{\gamma,Z}(x)$ are the following quark currents
\be
J_\nu^\gamma = \sum Q_q \bar{q}\gamma_\mu q~, \qquad 
J_\nu^Z = \frac{1}{2} \sum \bar{q}\gamma_\mu \left( g_V^q + g_A^q \gamma_5 \right) q~. \\
\ee
Taking into account that $J_\nu^Z = J_\nu^3 - s_W^2 J_\nu^\gamma$, where $J_\nu^3 = \sum T_3^q \bar{q}_L\gamma_\mu q_L$, we can express
$\Pi_{\mu\nu}^{Z\gamma}$ as a linear combination of $\Pi_{\gamma\gamma}(q^2)$ and  $\Pi_{3\gamma}(q^2)$. 
Since $J_\nu^\gamma$ has no axial component, only the vector part of $J_\nu^{Z(3)}$ contributes to these correlators. 

The QCD vacuum-polarization functions $\Pi_{\gamma\gamma}(q^2)$ and  $\Pi_{3\gamma}(q^2)$ have been extensively discussed in the literature
in the context of electroweak precision observables (see e.g. Ref.~\cite{Jegerlehner:1985gq}).
These functions can be reliably calculated using perturbative QCD only for $q^2 \gg \Lambda_{QCD}^2$ and sufficiently far from quark-antiquark production thresholds and narrow resonances. 

As pointed out first in Ref.~\cite{CabibboGatto},   $\Pi_{\gamma\gamma}$ can be extracted from $\sigma(e^+e^-\to \mbox{hadrons})$ data 
for any value of $q^2 >  4 m_e^2$  using dispersion relations (based only on causality and unitarity of the $S$-matrix):
\ba
\Pi_{\gamma\gamma}(q^2) - \Pi_{\gamma\gamma}(0)
= \frac{q^2}{\pi} \int_0^\infty ds \frac{\rm{Im} \Pi_{\gamma\gamma}(s)}{s(s-q^2-i\epsilon)}
= \frac{q^2}{12\pi^2} \int_0^\infty ds \frac{R(s)}{s(s-q^2-i\epsilon)}~,
\ea
where
\ba
R(s) \equiv \frac{\sigma(e^+e^-\to \mbox{hadrons})}{\sigma_0(e^+e^-\to \mu^+\mu^-)}
\ea
and $\sigma_0$ denotes the tree-level $e^+e^-\to \mu^+\mu^-$ cross-section. 

To extract $\Pi_{3\gamma}(q^2)$ from data 
we need some extra  theoretical assumptions. In the limit of exact $SU(3)$ symmetry for the light flavors, and 
taking into account that the OZI-rule is satisfied to good accuracy
for the heavy flavors, we can write~\cite{Jegerlehner:1985gq}
\ba
\Pi_{3\gamma}(q^2) \approx \frac{1}{2}\Pi_{\gamma\gamma}^{uds}(q^2) + \frac{3}{8}\Pi_{\gamma\gamma}^c(q^2) + \frac{3}{4}\Pi_{\gamma\gamma}^b(q^2)~,
\ea
which implies 
\be
\Pi_{Z\gamma} (q^2) \approx \left(\frac{1}{2}-s_W^2\right)\Pi_{\gamma\gamma}^{uds} (q^2)
+ \left(\frac{3}{8}-s_W^2\right)\Pi_{\gamma\gamma}^c(q^2) + \left(\frac{3}{4}-s_W^2\right)\Pi_{\gamma\gamma}^b(q^2)~.
\ee

In the following we are interested in estimating $\Pi_{Z\gamma} (q^2)$  for $q^2 >  (2~{\rm GeV})^2$. In this 
region  $\Pi_{\gamma\gamma}^{uds}(q^2)$ can be reliably estimated in perturbation theory, while the contribution
to $R(s)$ due to $c\bar c$ and $b\bar b$ narrow resonances is well described by a sum of narrow Breit-Wigner terms. 
Neglecting the smooth contribution of open heavy flavor production, we can write 
\ba
\label{eq:resonance}
\Pi^q_{Z\gamma} (s) = \frac{1}{2} \sum_i g_V^q Q_q \frac{s f_{V_i}^2}{m_i^2(m_{V_i}^2-s-i\Gamma_{V_i} m_{V_i})}~, \qquad q=c,b~,
\ea
where the sum runs over hadronic $q\bar q$ resonances with $J^{CP}=1^{--}$,\footnote{In principle, $J^{CP}=1^{++}$ states contribute to the $h\to Z\ell^+\ell^-$ decay via the $\Pi_{ZZ}(q^2)$ correlator. However, the later gives a contribution 
that is $q^2/m_Z^2$ suppressed compared to $\Pi_{Z\gamma}(q^2)$ and thus can be safely neglected.} whose  
decay constants, $f_{V_i}$, are defined by
\ba
\langle 0 | \bar{q} \gamma_\mu q | V_i (p, \epsilon) \rangle  = f_{V_i} m_{V_i} \epsilon_\mu~.  
\ea

\subsection{Modifications of the $q^2$ spectrum due to narrow resonances}
\label{sect:charmonium}

\begin{figure}[p]
\centering
%
\includegraphics[width=0.45\textwidth]{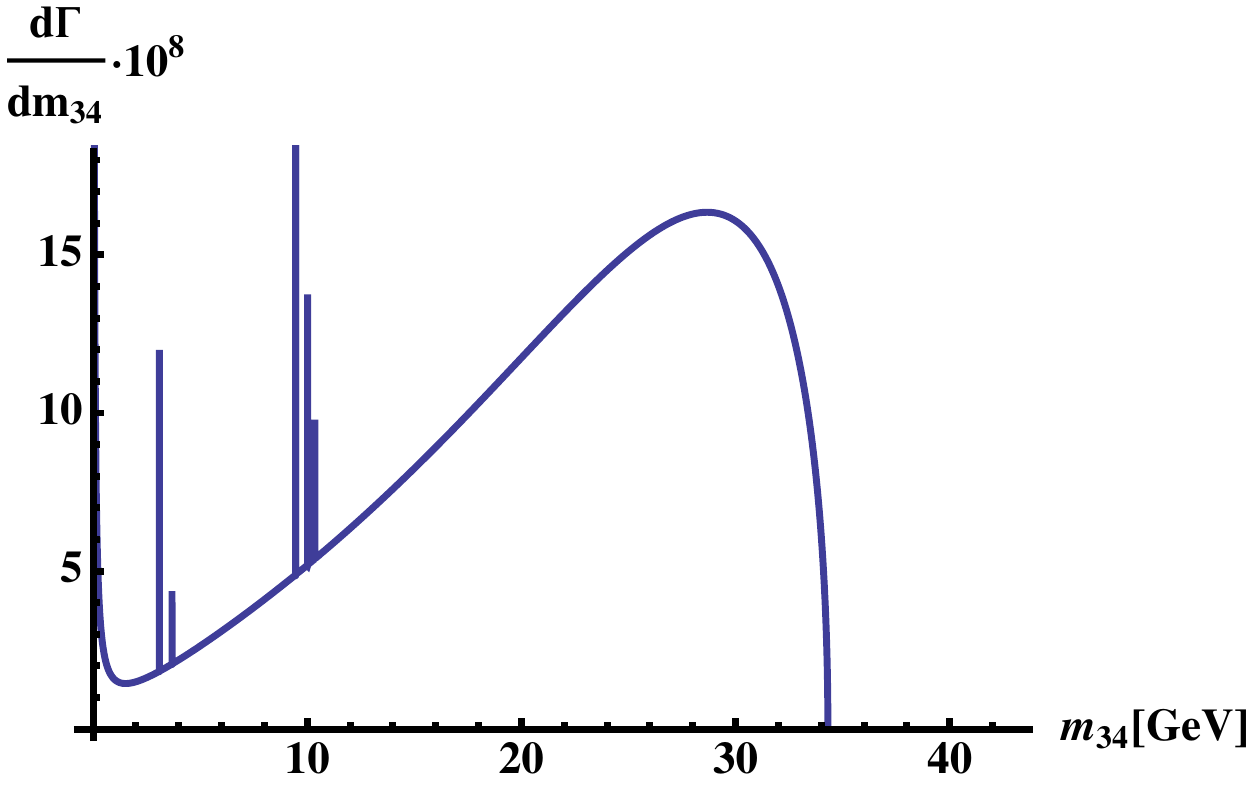}
\hspace{0.2in}
\includegraphics[width=0.45\textwidth]{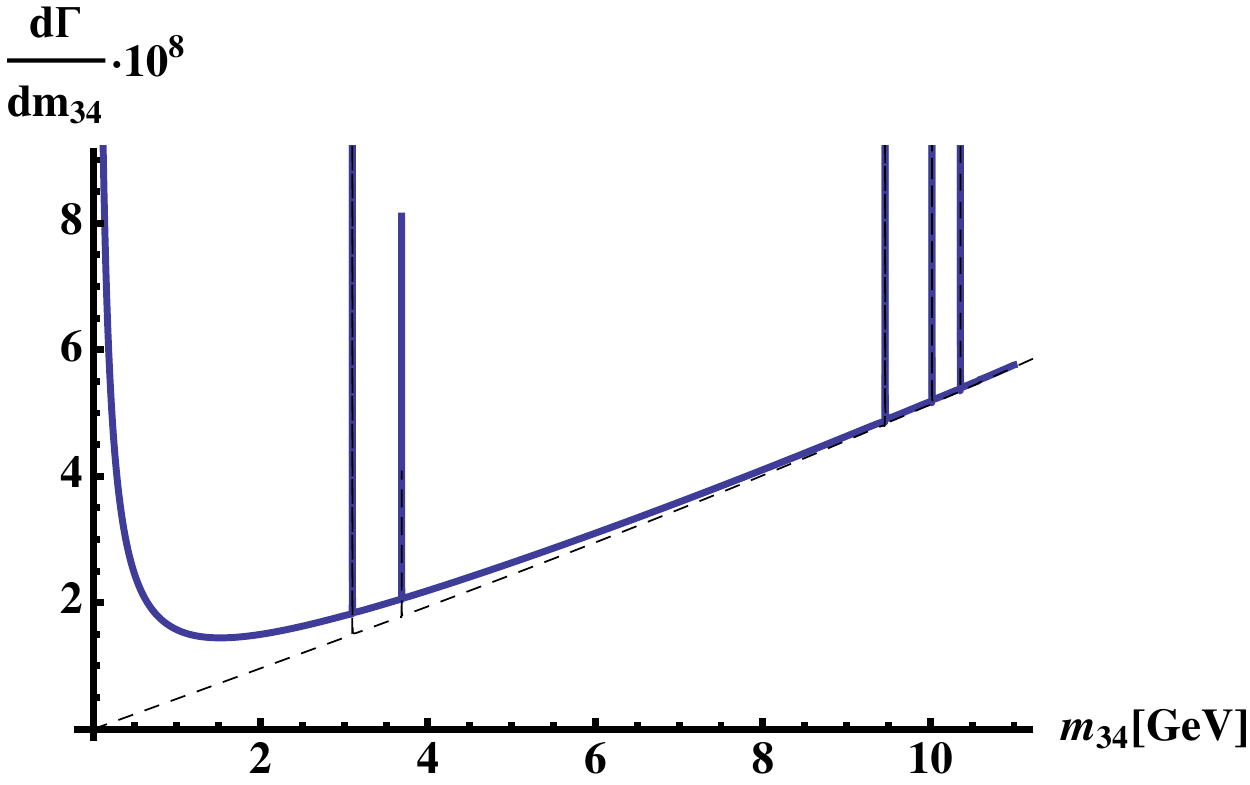}
\includegraphics[width=0.45\textwidth]{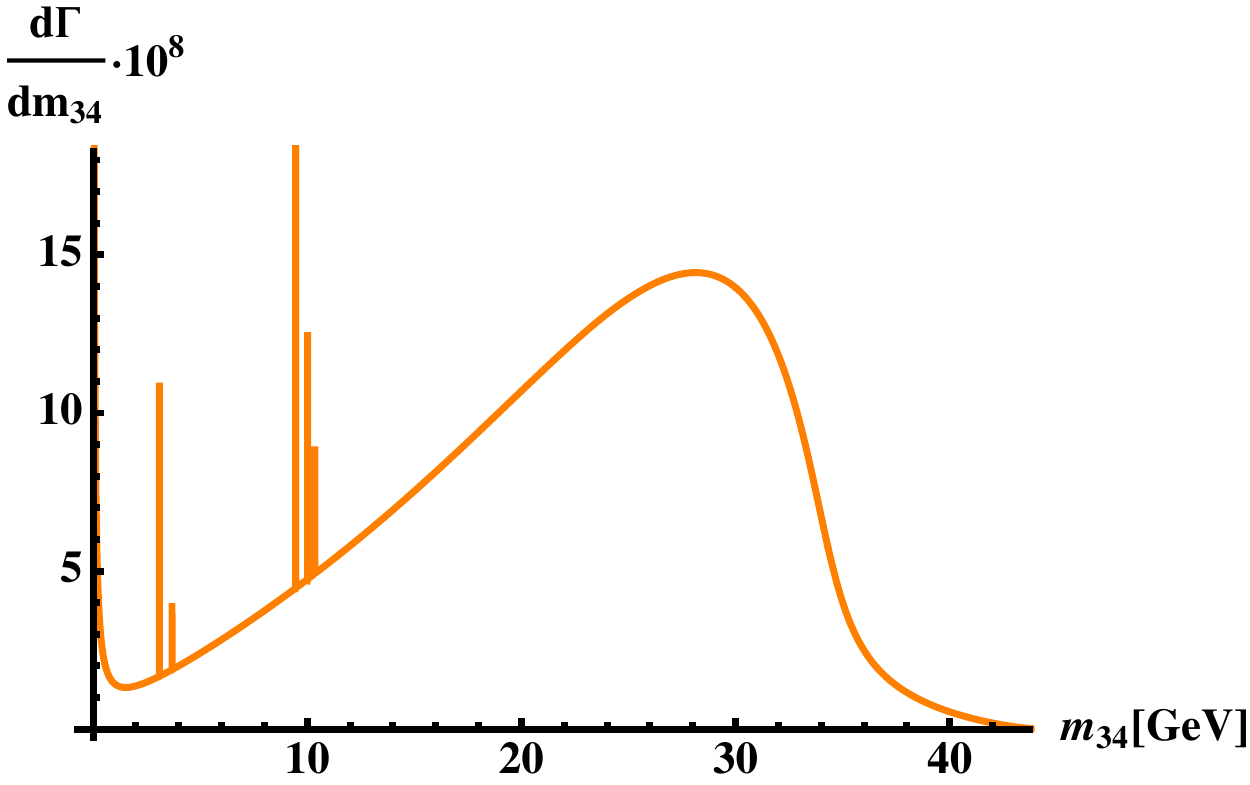}
\hspace{0.2in}
\includegraphics[width=0.45\textwidth]{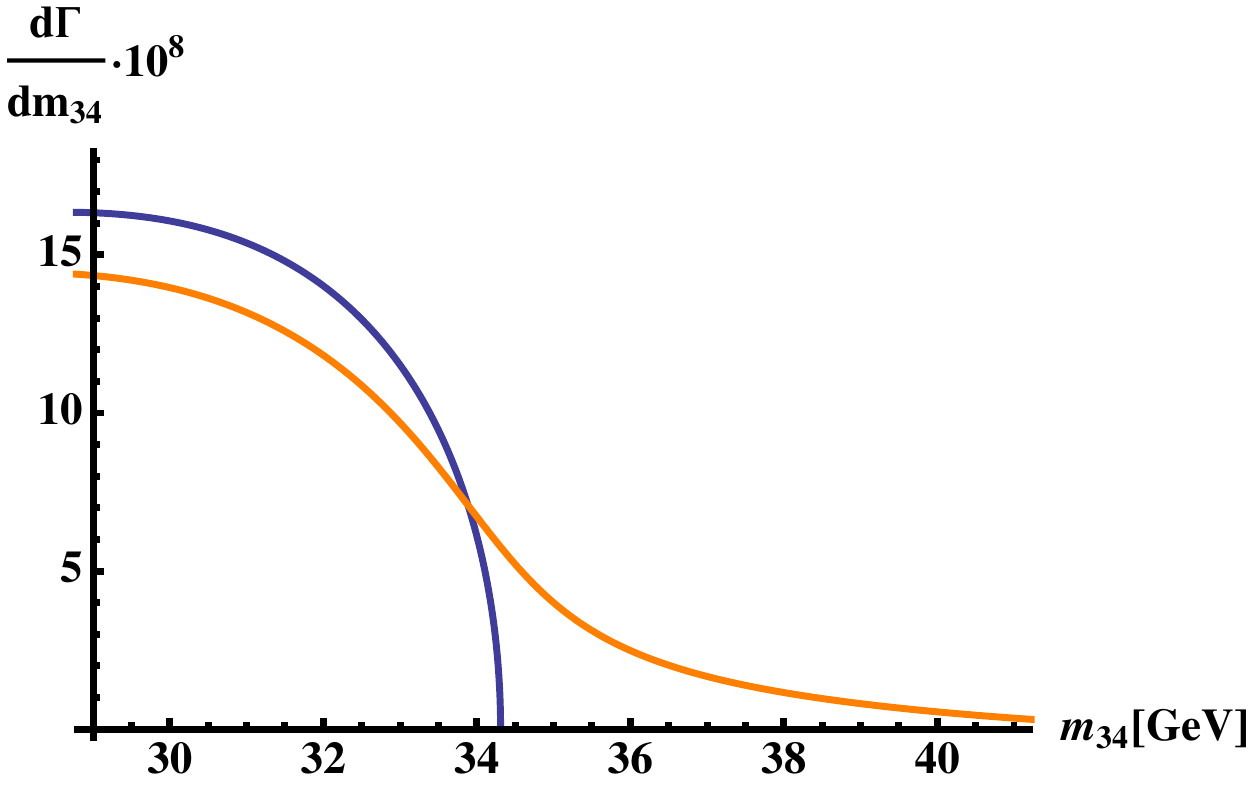}
\caption{First row: Spectrum including the effects from hadronic resonances and the $h\to Z\gamma^*$ amplitude,
with a zoom in the low-$m_{34}$ region (the dashed line in the right plot is obtained neglecting the $h\to Z\gamma^*$ amplitude).
Third row: spectrum after the $m_{12}$ smearing due to the off-shelleness of the $Z$ boson ($|m_{12} - m_Z| \leq 10$ GeV), with a zoom in the large-$m_{34}$ region (the blue line in the right plot is obtained without smearing).} 
\label{fig:GammaCorrected}
\end{figure}

The NLO contributions due to $Z-\gamma$ mixing can be included in the $h\to Z\ell^+\ell^-$ decay 
distribution via the following straightforward modification of Eq.~\eqref{eq:gamma0}:
\ba
\left[ (g_R^\ell)^2 + (g_L^\ell)^2 \right]  & \longrightarrow &  \frac{1}{2} \left[ 
(g_A^\ell)^2 +  \left| g^\ell_V +2e^2 \Pi_{Z\gamma}(q^2) \right|^2  \right] ~, 
\ea
Evaluating $\Pi_{Z\gamma}(q^2)$ by means of Eq.~\eqref{eq:resonance}  we are thus able to include the distortions 
of the $q^2$ spectrum due to hadronic resonances. Taking into account all the $J^{CP}=1^{--}$ resonances listed 
in the PDG we derive to the $q^2$ distributions shown in Fig.~\ref{fig:GammaCorrected}, where the additional correction 
due to the  $h\to Z\gamma^*$ amplitude in Eq.~\eqref{eq:gamma1} has also been included.
As expected, the only visible 
peaks above 2 GeV are those induced by the narrow $J/\Psi(nS)$ and $\Upsilon(nS)$ resonances that cannot decay into open charm 
and open bottom states, respectively. Their complete list is reported in Table~\ref{tab:Vrates}.\footnote{The $\sim 20\%$ difference between the value of  
$\BR(h \to Z J/\psi)$ in Table~\ref{tab:Vrates}  and the original prediction in Ref.~\cite{Isidori:2013cla} is due to updated (more accurate) 
numerical inputs.}

The plots shown in the first row of Fig.~\ref{fig:GammaCorrected} assume an exactly on-shell $Z$, whereas in the third row we take into account the full $h\to 4\ell$ decay, and impose the following cut on $m_{12}$ around the $Z$ mass: $|m_{12} - m_Z| \leq 10$ GeV~. As can be noted, this imply a significant smearing near the end point of the $m_{34}$ distribution, but has almost no impact in the shape of the resonance region.

\begin{figure}[p]
\centering
\includegraphics[width=0.45\textwidth]{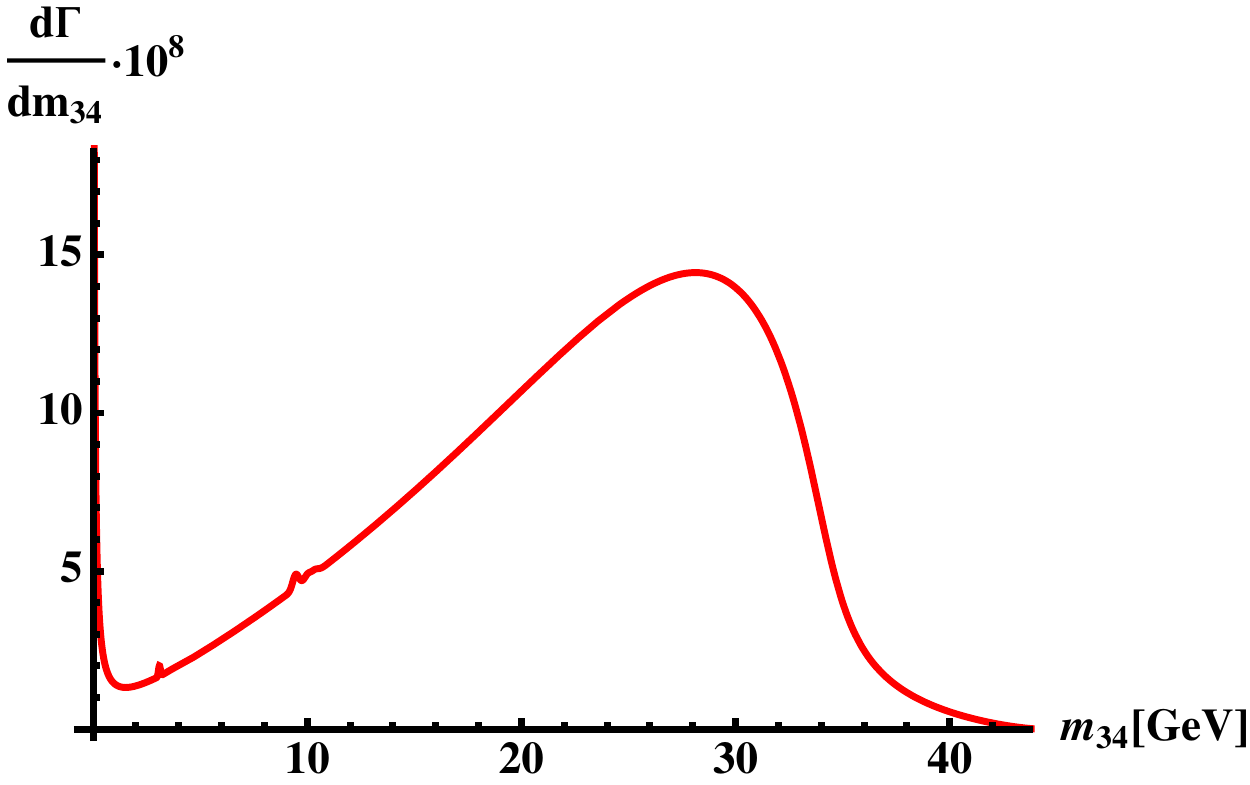}
\hspace{0.2in}
\includegraphics[width=0.45\textwidth]{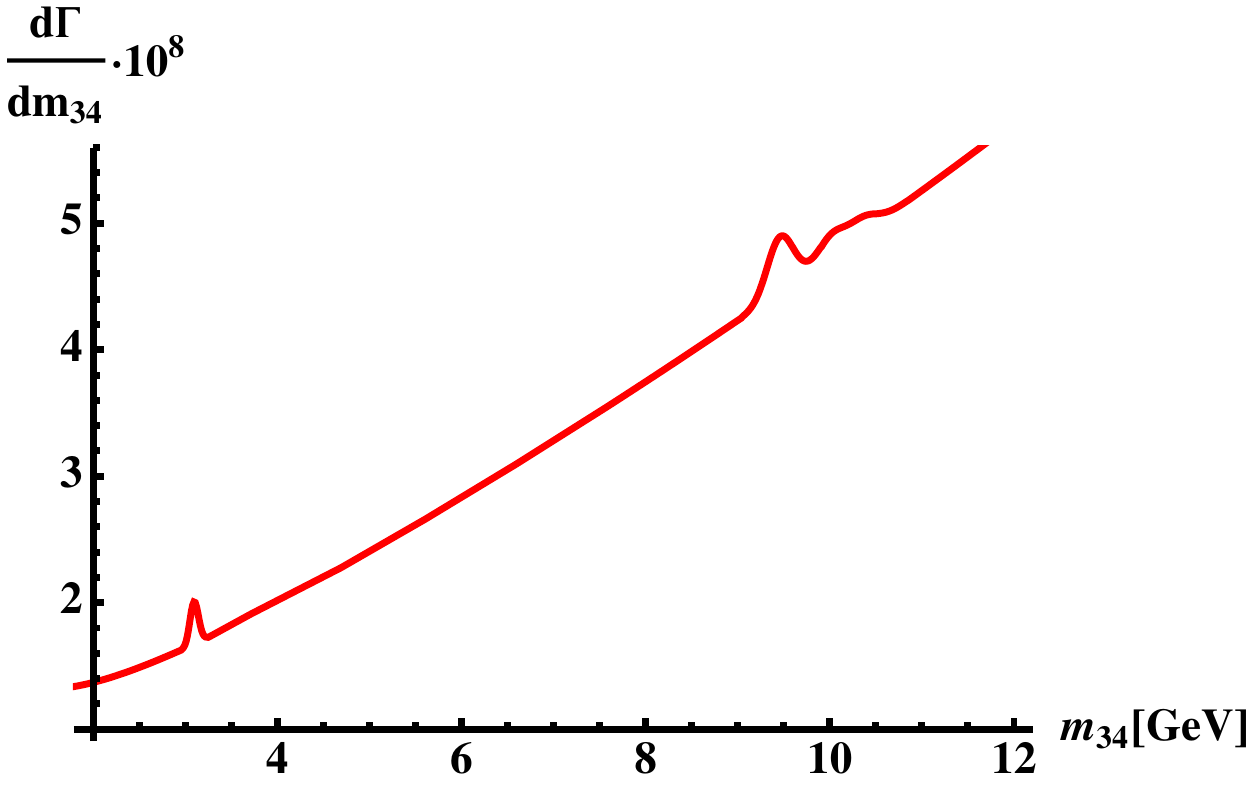}
\includegraphics[width=0.45\textwidth]{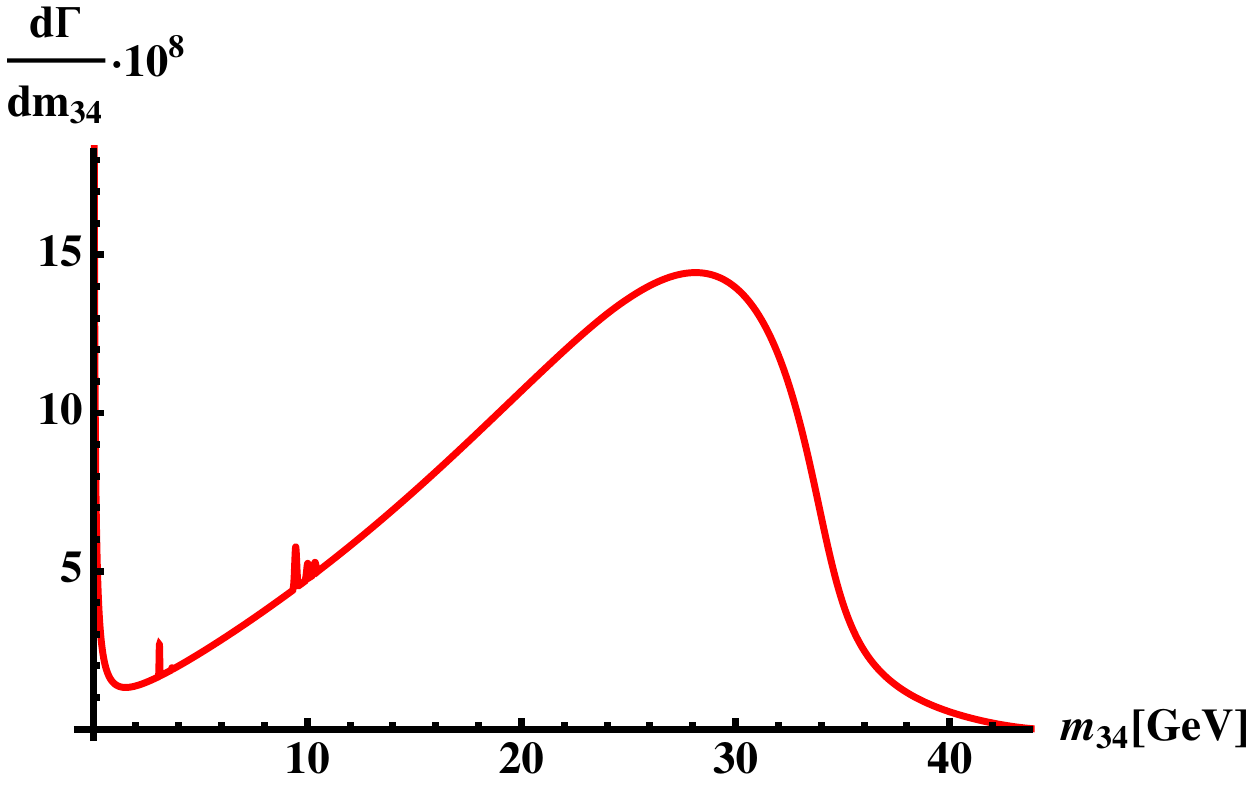}
\hspace{0.2in}
\includegraphics[width=0.45\textwidth]{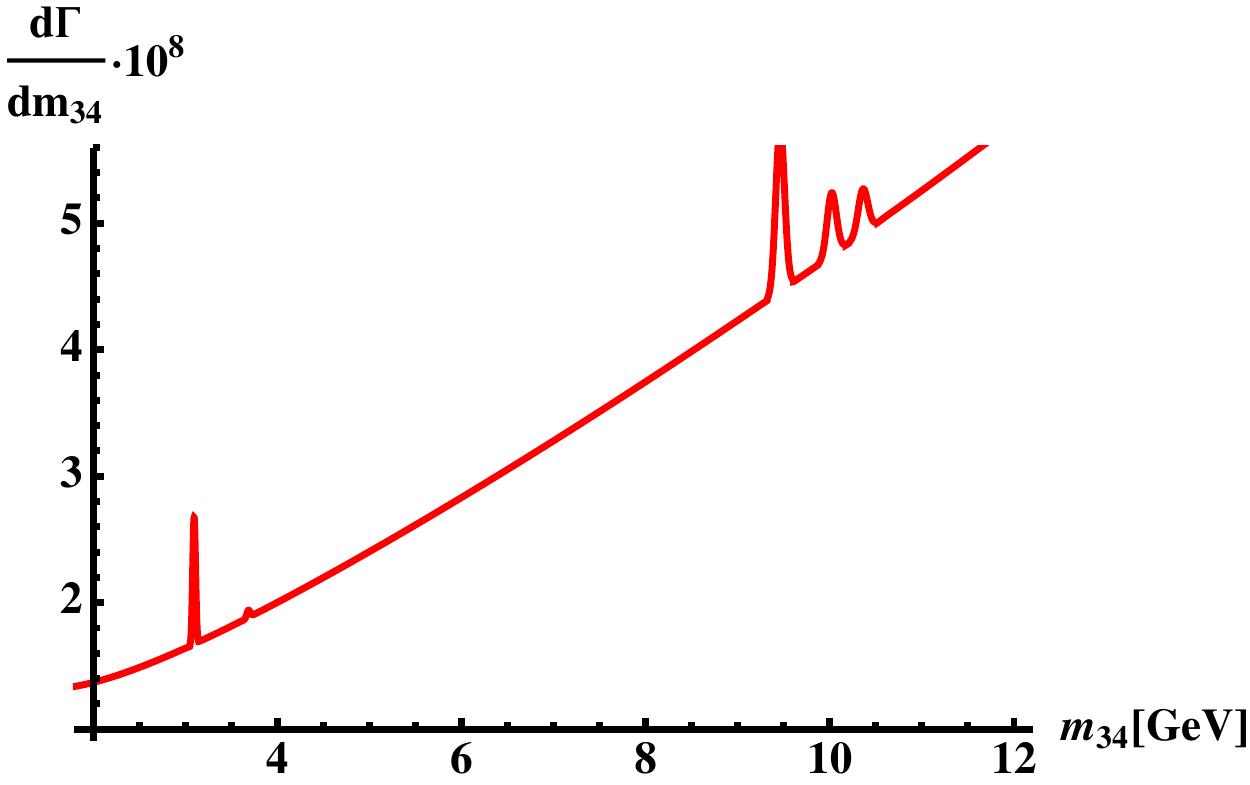}
\caption{Spectrum convoluted with an idealized experimental smearing in $m_{34}$, with a zoom in the resonance region (right plots). The first (second) row assumes an experimental resolution of $\sigma_{m_{\ell\ell}}= 1.5\%\times m_{\ell\ell}$ ($\sigma_{m_{\ell\ell}}= 0.5\%\times m_{\ell\ell}$).} 
\label{fig:smearing}
\end{figure}

\begin{table}[t]
\begin{center}
\begin{tabular}{|c|c|c|c|c|} \hline
State & $m_{V_i}$[GeV] & $f_{V_i} $[MeV] & $\BR(h \to ZV_i) $ &  \raisebox{1pt}{$\Delta [d\Gamma(h \to Z\ell\ell)/dm_{34}]_{[1~{\rm GeV~bin}]}$}  \\ \hline
$J/\psi$(1S) & 3.10  &    405  & $1.7 \times 10^{-6}$  & 2.6\% \\
$J/\psi$(2S) & 3.69  &    290 & $ 8.6 \times 10^{-7}$ & 0.2\%\\ \hline
$\Upsilon$(1S) &9.46 & 680  & $1.6 \times 10^{-5}$  & 3.1\% \\
$\Upsilon$(2S) &10.02  &  485  & $ 8.2 \times 10^{-6}$ & 1.2\% \\ 
$\Upsilon$(3S) & 10.36  & 420  & $ 6.2 \times 10^{-6}$ & 0.9\% \\ 
\hline
\end{tabular}
\caption{\label{tab:Vrates}
List of narrow $c\bar c$ and $b\bar b$ narrow resonances giving rise to sizable modifications of the $d\Gamma(h \to Z\ell\ell)/dq^2$ spectrum. In the last column we report the relative modification of the spectrum assuming the following $m_{34}$ bin: $[m_{V_i} -\Delta/2, m_{V_i} +\Delta/2]$, with $\Delta =1$~GeV.}
\end{center}
\end{table}

In the limit $\Gamma_{V_i}/m_{V_i} \to 0$, which is a very good approximation for the narrow resonances in  Table~\ref{tab:Vrates}, 
the contribution of the narrow states adds incoherently to the leading (perturbative) contribution in the $h\to Z\ell^+\ell^-$  spectrum. The incoherent 
contribution of each resonance is given by
\ba
\Gamma(h\to Z V_i \to Z \ell^+\ell^-) \approx \Gamma(h\to Z V_i ) \times \BR(V_i \to \ell^+\ell^-)~,
\ea
where
\ba
\BR(V_i \to \ell^+\ell^-) & = & \frac{4\pi Q_q^2}{3} \frac{\alpha^2 f_{V_i}^2}{m_{V_i} \Gamma_{V_i}} \left[ 1 +O\left(m^2_\ell/{m_{V_i}^2}\right)\right]  ~,  \\
\Gamma(h\to Z V_i) &=& \frac{1}{16\pi}  \frac{m_h^3   }{v^4} (g^q_V f_{V_i})^2 
 \frac{\sqrt{\lambda\left( 1,\hrho, \hat \epsilon \right)}}{(1-\hat \epsilon/\hrho)^2} \left[ (1-\hrho)^2 \left(1-\frac{\hat \epsilon}{1-\hrho} \right)^2 + 8 \hrho\hat \epsilon \right]
 \nonumber \\
& = &  \frac{  (1-\hrho)^3 }{16\pi}  \frac{m_h^3   }{v^4} (g^q_V f_{V_i})^2  R~,
\ea
$\hat \epsilon=m_{V_i}^2/m_h^2$ and $R = 1 + O(\hat \epsilon)$.\footnote{Expanding up to linear order in $\hat \epsilon$
one gets  $R = 1 +  \frac{11\hrho^2-7\hrho+2}{\hrho(1-\hrho)^2} \hat \epsilon + {\cal O}(\hat \epsilon^2) \approx 1+12\hat \epsilon$.}
Using $\Gamma_h (m_h=125.5~{\rm GeV})\approx 4.07$ MeV we find the relative rates reported in Table~\ref{tab:Vrates}. As  shown in the last column of Table~\ref{tab:Vrates}, the hadronic resonances cause at most a $\sim$3\% modification in a 1 GeV wide bin of $m_{34}$, but the relative impact would raise to $\sim$30\% for a $100$~MeV wide bin, assuming an experimental $m_{34}$ resolution significantly smaller than the bin width. An effect of this size could possibly be measurable at the LHC in the high-luminosity phase.

In order to take into account the finite experimental resolution we include a simple gaussian smearing in the measurement of $m_{34}$, obtaining the results shown in Fig.~\ref{fig:smearing}. In particular, we assume $\sigma_{m_{\ell\ell}}= 1.5\%\times m_{\ell\ell}$ (upper panel) and $0.5\%\times m_{\ell\ell}$ (lower panel).  As can be seen, the narrow Breit-Wigner peaks become approximate (sometimes overlapping) gaussian  curves of $\sigma_{m_{\ell\ell}}^2$ variance. 


\section{NP contributions and connections with $(g-2)_\mu$.}
\label{sect:NP}

The anomalous magnetic moment of the muon, $a_\mu = (g-2)_\mu /2$, is a very sensitive probe of physics beyond the SM. At present, 
$a_\mu$ is also one of the very few low-energy observables exhibiting a significant deviation 
between data~\cite{Bennett:2006fi} and SM prediction~\cite{Jegerlehner:2009ry}:
\be
\Delta a_\mu \equiv a_\mu^{\rm exp} - a_\mu^{\rm th} =  (2.9 \pm 0.9)  \times 10^{-9}~.
\label{eq:amuexp}
\ee
In the following we will assume that this discrepancy is due to NP, and explore the possible implications of this 
assumption in $h\to 4\ell$  decays.

On general grounds, it is natural to expect a connection between NP contributions to $a_\mu$ and possible deviations from the 
SM in $h$ decays of the type $ h\to \mu^+\mu^- + X_{\rm em}$, where $X_{\rm em}$ is either a photon or a state coupled to the 
electromagnetic current ($X_{\rm em} = \gamma$, $e^+e^-$, $\mu^+\mu^-$, \ldots).  
This connection is manifest by looking at the $SU(2)\times U(1)_Y$ invariant effective operator 
describing the $a_\mu$ anomaly in generic extensions of the SM (see below).
However, we should distinguish two main scenarios: i) NP models where the new particles generating $\Delta a_\mu$
have a mass above (or around) the electroweak scale; ii) NP models where the new particles generating $\Delta a_\mu$
are light and can be produced on-shell in $h$ decays. 

\subsection{NP above the electroweak scale}
If the SM is the low-energy limit of a theory with new states above the electroweak (EW) scale respecting the 
$SU(2)\times U(1)_Y$ symmetry, all NP effects can be parametrized by higher-dimensional $SU(2)\times U(1)_Y$ invariant 
operators. After EW symmetry breaking, the only combination of effective operators contributing at $a_\mu$ at the tree level is
\be
\cL_{\rm EFT} =  \frac{c_0}{\Lambda^2}   ~ \bar L^{(\mu)}_L \sigma^{\mu\nu} \mu_R F_{\mu\nu}H\,+\mbox{h.c.}\,, ~~~
\left.  H \right|_{\rm unit.\,gauge} = \frac{1}{\sqrt{2}} \left(\!\!\begin{array}{c} 0 \\ h+ v \end{array}\!\!\right)\,, 
~~~ L^{(\mu)}_L = \left(\!\!\!\begin{array}{c} \nu_{L}^{(\mu)} \\ \mu_L \end{array}\!\!\!\right)~,
\label{eq:LEFT}
\ee
where $F_{\mu\nu}$ it the electromagnetic field tensor. 
This operator affects at the tree level also $ h\to \mu^+\mu^- + X_{\rm em}$ decays, through the $h\to \mu\mu \gamma$
amplitude. The NP contribution to $a_\mu$ is
\ba
\Delta a_\mu 
= - \frac{c_0}{\Lambda^2}  \frac{4 m_\mu v}{\sqrt{2} e}~
\approx~Ê-5 \times 10^{-9}\,\frac{c_0}{y_\mu}   \left( \frac{5~\mbox{TeV}}{ \Lambda} \right)^2~,
\ea
where we see that TeV-physics with a MFV-like coupling (here $y_\mu= \sqrt{2}m_\mu/v$)
can naturally generate the current anomaly. 
However, such a value for $c_0/\Lambda^2$ is by far too small to generate any observable effect in $h$ decays.
Taking into account the interference with the tree-level $h\to \mu^+\mu^-\gamma$ 
SM amplitude we find
\ba
\Delta \Gamma(h \to \mu^+\mu^- \gamma )^{(g-2)}_{\rm EFT} &=& - \frac{e^2 m_h^3  \Delta a_\mu }{128 \pi^3 v^2} 
+ \frac{e^2 m_h^5  \left(\Delta a_\mu\right)^2}{12 (8 \pi)^3 m_\mu^2 v^2}  \approx - 2\times10^{-12} \mbox{ GeV}~, 
\ea
where the result is dominated by the interference term. This imply 
$\Delta \BR (h \to \mu^+\mu^- \gamma )^{(g-2)}_{\rm EFT}  = O(10^{-10})$, namely a 
$O(10^{-4})$ correction with respect to $\BR (h \to \mu^+\mu^- \gamma )_{\rm SM}$,
that is beyond any realistic detection.  Indeed the present experimental 
limit on $\BR (h \to \mu^+\mu^- \gamma )$ is about 10 times above the SM prediction~\cite{CMSmmg}. 


We thus conclude that if particles above the weak scale are at the origin of the $(g-2)_\mu$ anomaly, we should not expect any directly related visible impact in $h$ decays. 
Needless to say, if the heavy particles are within the LHC reach, they could be directly produced in $pp$ collisions. However, 
the connection with $\Delta a_\mu$ is more model-dependent in this case (see Ref.~\cite{Freitas:2014pua} for a recent attempt to analyse
in general terms  the connection between $\Delta a_\mu$  and new states within the LHC reach).


\subsection{Light scalar}
\label{eq:scalar}
A somehow orthogonal explanation of the $(g-2)_\mu$ anomaly is obtained assuming the existence of new light states, with mass 
$m_\mu \ll m_{\rm NP}  \ll m_h$. In this framework the contribution to $\Delta a_\mu$, or a non-vanishing effective coupling for the 
operator in Eq.~(\ref{eq:LEFT}),  is generated at the one-loop level, 
while NP can have a significantly larger impact in $h$ decays with the 
direct (tree-level) production of the new states.
Here we study a prototype case in this category, extending the SM with a single, 
$SU(2)\times U(1)_Y$ invariant, scalar field $\phi$.
We assume that the effective interaction of $\phi$ to muons is generated by the exchange of additional heavy (TeV-scale) particles, 
resulting into the following effective Lagrangian:
\be
\cL^{(1)}  =   \cL^{(\phi)}_{\rm kin} +\left(  \frac{\cmu} {\Lambda} \bar L^{(\mu)}_L \mu_R H \phi~ +{\rm h.c.} \right)~, \qquad 
\cL^{(\phi)}_{\rm kin} =  \frac{1}{2} \partial_\mu \phi\partial^\mu\phi -\frac{1}{2} m_\phi^2 \phi^2~.
\ee

The one-loop contribution   to $a_\mu$ generated by $\cL^{(1)}$ is 
\ba
\Delta a_\mu = \frac{|\cmu|^2 }{96\pi^2}  \frac{ v^2 }{\Lambda^2} \frac{m_\mu^2 }{ m_\phi^2}
~\approx~ 6.4 \times 10^{-9} ~ |\cmu|^2~ \left( \frac{1~\rm TeV}{ \Lambda} \right)^2    \left( \frac{10~\rm GeV}{ m_\phi} \right)^2~,
\ea
which can easily accommodate the current experimental anomaly for $m_\phi =O(10~{\rm GeV})$ and $\cmu =O(1)$.
Note that the sign of the contribution to $\Delta a_\mu$ is necessarily positive, 
in agreement with the experimental result in Eq.~(\ref{eq:amuexp}).

The same interaction leads to a non-standard (tree-level) $h \to \mu^+\mu^- \phi$ decay with the following differential rate
\ba
 \frac{d\Gamma(h\to \mu\mu\phi)}{dm_{12}} = \frac{|\cmu|^2}{128 \pi^3 m_h^3 \Lambda^2} ~m^3_{12}(m_h^2 - m^2_{12})~,
 \label{eq:hmmphi}
\ea
where $m_{12}$ is the invariant mass of the muon-antimuon pair. 
The total branching ratio is 
\be
\BR(h \to \mu^+\mu^- \phi ) 
= \frac{|\cmu |^2  m^3_h }{1536 \pi^3  \Lambda^2 \Gamma_h}
~\approx~4.8 \times 10^{-3} 
  \left(\frac{ \Delta a_\mu }{ 2.9\times 10^{-9} } \right) \left( \frac{ m_\phi}{10~\rm GeV} \right)^2~,
  \label{eq:BRhmm}
\ee
where we have exchanged $\cmu$ by $\Delta a_\mu$, leaving the mass of the scalar $m_\phi$ as the only free parameter. 
The new scalar certainly decay into a pair of muons (by construction we assume $m_\phi >2 m_\mu$) 
but may also have other decay modes  (both SM particles, such as neutrinos, or other light exotic states). We can thus write 
\ba
\Gamma_\phi > \Gamma (\phi\to\mu\mu) 
= \frac{|\cmu|^2  v^2 m_\phi }{16 \pi  \Lambda^2 }
\approx 
( 5.9 \mbox{ MeV}) \times \left(\frac{ \Delta a_\mu }{ 2.9\times 10^{-9} } \right) \left( \frac{ m_\phi}{10~\rm GeV} \right)^3~,
\ea
from which we conclude that $\phi$ is not long-lived. The $h\to \mu\mu\phi$ channel could be detected 
only as an additional contribution to the $4 \mu$ final state,
as discussed below.

\subsubsection*{The  $h \to \mu\mu\phi  \to 4 \mu$ decay} 
Using the above expressions we find
\ba
\frac{ \BR( h \to 4 \mu)_{(\phi)} }{ \BR(h \to 4 \mu)_{\rm SM} } & \approx & 150  \left(\frac{ \Delta a_\mu }{ 2.9\times 10^{-9} } \right) \left( \frac{ m_\phi}{10~\rm GeV} \right)^2 \BR(\phi\to \mu^+\mu^-)~,
\label{eq:4mu_phi0}
\ea
where we see that the NP effect would exceed the SM rate, unless $\BR(\phi\to \mu^+\mu^-)\ll1$ or $m_\phi < 1$~GeV. 

The distribution in Eq.~(\ref{eq:hmmphi}) has a kinematical peak at $m_{12} = \sqrt{3/5}~m_h \approx 97$~GeV, 
which is remarkably close to $m_Z$. As a result, when organizing the dimuon pairs in low
and high masses, the resulting double-differential spectrum 
$d^2 \Gamma(h\to 4\mu )/dm_{12} dm_{34}$
is quite similar to the one expected in the SM 
for the quarkonium resonances (with the charmonium mass replaced by $m_\phi$). 
Indeed, the Z-pole cut $|m_{12} - m_Z|<\Delta$ performed by current experimental analysis will be passed by a significant fraction of the events produced through the light scalar interaction. Namely
\ba
\frac{ \BR (h\to (\mu\mu)_Z (\mu\mu)_\phi) }{ \BR(h \to 4 \mu)_{\rm SM} }  
= f\times \frac{\BR( h \to 4 \mu)_{(\phi)}}{\BR(h \to 4 \mu)_{\rm SM}}~,
\qquad
 f \approx ~24\,\rho^{3/2} (1-\rho) \frac{\Delta}{M_h}~,
\label{eq:4mu_phi1}
\ea
which implies $f\approx 0.35 \,(0.70)$ for $\Delta=10 \,(20)$ GeV. 
Thus, an experimental limit of 50\% on this ratio\footnote{A 90\%CL limit close 50\% is what we deduce from current data~\cite{Aad:2013wqa,Chatrchyan:2013mxa}, for $m_\phi \gtrsim 12$~GeV.}
would imply
\ba
\left(\frac{ \Delta a_\mu }{ 2.9\times 10^{-9} } \right) \left( \frac{ m_\phi}{10~\rm GeV} \right)^2 \BR(\phi\to \mu^+\mu^-)~ < 0.003/f~.
\label{eq:bound50}
\ea
However, we stress that if $m_\phi  \lesssim 12$ GeV, a large fraction of this hypothetical exotic decay is not selected  applying the cuts presently applied by ATLAS and CMS in their $h\to 4 \mu$ analyses.

As in the charmonium case, the peak at $m_{34}=m_{\phi}$ in the differential decay rate $d \Gamma(h\to 4\mu )/dm_{34}$ represents a much better observable to search for this exotic interaction. 
In fact, assuming $m_\phi$ and  $\BR(\phi\to \mu^+\mu^-)$ saturate the bound in Eq.~(\ref{eq:bound50}), we would find $\BR (h\to (\mu\mu)_Z (\mu\mu)_\phi) \sim 900$ times larger than the contribution due to the $\Upsilon(1S)$ meson. 
In such case, the NP effect in a 1 GeV wide bin around the $\phi$ mass would be $\sim 30$ times larger than the SM: an effect that is likely to be 
already ruled out by  present data.

\subsubsection*{Other $h$ decay modes} 

If the $\phi$ has a non-vanishing decay into $e^+e^-$ pairs, the $h \to \mu\mu\phi$ amplitude could have  
a non-negligible impact also in 
$h \to 2 \mu 2e $ decays. The relative impact, compared to the SM, can trivially be obtained replacing $\BR(\phi\to \mu^+\mu^-)$ with 
$\BR(\phi\to e^+e^-)$   in Eq.~(\ref{eq:4mu_phi0}). However, we stress that while a non-vanishing 
$\BR(\phi\to \mu^+\mu^-)$ is guaranteed by the contribution of $\phi$ to the $(g-2)_\mu$ anomaly, in principle
$\BR(\phi\to e^+e^-)$ could be very suppressed.

A ``true'' $h \to Z \phi $ decay could occur if we add the following interaction term, 
\be
\Delta \cL^{(1)}  =     \frac{\ch} {2\Lambda} \left( i  H^\dagger D_\mu H \partial^\mu  \phi~+~{\rm h.c.}\right)~.
\ee
This interaction induces a quadratically divergent contribution to the $Z$-boson mass: $\Delta m_Z^2 / m_Z^2  \approx  \ch^2/(32\pi^2)$.
Imposing $\Delta m_Z^2/m_Z^2 < 5\times 10^{-4}$ from electroweak precision observables leads to $|\ch| <  0.4$. 
Once this bounds is satisfied and $|\cmu|$ satisfies the $\Delta a_\mu$ bound, all other constraints (in particular from muon lifetime
and Higgs mass) are satisfied. The new exotic channel has the following relative rate
\be
\BR(h \to Z\phi) 
= \frac{|\ch|^2  m^3_h }{64\pi \Lambda^2 \Gamma_h} \lambda( \hrho, \hat m^2_\phi)^3~
\approx ~ 0.14 \left|\frac{\ch}{0.4} \right|^2 \left(\frac{1~\mbox{TeV}}{\Lambda} \right)^2 ~,
\ee
where $\hat m^2_\phi =   m_\phi^2 / m_h^2 $, which could exceed $\BR(h \to \mu\mu\phi)$  in Eq.~(\ref{eq:BRhmm}).
Considering only this exotic channel we find 
\ba
\frac{ \BR[ h \to (2\ell)_Z (2 \mu)_{(\phi)} ] }{ \BR(h \to 2 \ell 2 \mu)_{\rm SM} } 
& \approx & 160  \left|\frac{ \ch }{ 0.4} \right|^2 \left( \frac{ 1~\rm TeV}{\Lambda} \right)^2 \BR(\phi\to \mu^+\mu^-)~,
\ea
which, similarly to Eq.~(\ref{eq:4mu_phi0}) and (\ref{eq:4mu_phi1}), 
can be used to obtain non-trivial constraints over the parameter space of the model. 
We stress that once again that 
the peak in the differential decay rate $d \Gamma(h\to 4\mu )/dm_{34}$ offers a better signal to background ratio than the total branching ratio.


We finally mention that in this framework is natural to expect a non-vanishing $h \to 2\phi$ decay
(e.g.~from the $d=4$ operator $H^\dagger H \phi^2$). However, the related 
$h \to 2\phi \to 4\mu$ spectrum is quite different from the SM one, and the corresponding 
effective coupling is unrelated to the $a_\mu$ anomaly. For these reasons, and  given 
this process has been extensively studied elsewhere (see Ref.~\cite{Curtin:2013fra} and references therein), 
we do not discuss it here.

\subsection{Light vector}

Following the renewed interest in NP models with light massive gauge fields~\cite{ArkaniHamed:2008qn}, 
the possibility to explain the $(g-2)_\mu$ anomaly by means of an exotic light vector particle
has been discussed in specific frameworks~\cite{Fayet:2007ua,Pospelov:2008zw,Carone}.

Adopting a general effective-theory approach, the leading ($d=4$) interactions of the exotic 
massive neutral vector $Z^\mu_d$ to muons can be parameterized as follows
\be
\cL^{(2)}_{\rm int}   =  -Z_d^\mu \left(  c_L \bar \mu_L \gamma_\mu \mu_L~ + c_R \bar \mu_R \gamma_\mu \mu_R \right)~,
\ee
where $Z_d^\mu$ is the mass eigenstate after electroweak symmetry breaking.
The one-loop contribution to $a_\mu$ expressed in terms of $m_{Z_d}$ and $c_{L,R}$  is:
\ba
\Delta a_\mu
= -\frac{1 }{12\pi^2} \frac{m_\mu^2 }{ m_{Z_d}^2} \left( c_R^2 + c_L^2 - 3c_R c_L \right) 
\approx  2.3\times 10^{-9} \left( \frac{10~\mbox{GeV}}{m_{Z_d}} \right)^2 \frac{ c_V^2 - 5 c_A^2}{0.1^2}~,
\ea
where $c_{V/A} = c_R \pm c_L$. The values of $c_{L,R}$ can be determined in specific models. For instance,  if the $Z_d$ interacts with SM fields only 
via a kinetic mixing of the form $\frac{1}{2}\frac{\epsilon}{\cos\theta_W}B_{\mu\nu}Z_d^{\mu\nu}$ 
(the so-called dark photon~\cite{ArkaniHamed:2008qn} hypothesis), 
one has $c_L=c_R=-e\,\epsilon$, up to ${\cal O}(m_{Z_d}^2 / m_{Z}^2)$ corrections.
In more general setups, the $Z_d$ field can also have a mass mixing with the $Z$ boson  of the form 
$-\epsilon_Z m_Z^2 Z_d^\mu Z_\mu$~\cite{Davoudiasl:2012ag}. Taking into account both forms of mixing, and 
allowing also non-vanishing charges for the muons under the Abelian group $U(1)_d$
associated to $Z_d$, we can write
\ba
c_L &=& -e\,\epsilon - \frac{g}{2c_W} (1-2s^2_W) \epsilon_Z + g_d Q_{\mu_L}^d~, \nonumber\\
c_R &=& -e\,\epsilon + \frac{g}{c_W} s^2_W \epsilon_Z + g_d Q_{\mu_R}^d~,  
\label{eq:cLcR}
\ea
up to ${\cal O}(m_{Z_d}^2 / m_{Z}^2)$ corrections.\footnote{~Here $g_d$ denotes the coupling to the $U(1)_d$ group
and $Q_{\mu_{L(R)}}^d$ the corresponding charges of $\mu_{L(R)}$.} The bounds on $\epsilon$ and $\epsilon_Z$ have been discussed 
in the literature (see e.g.~Ref.~\cite{Curtin:2013fra,Davoudiasl:2012ag} and references therein). Given the stringent bounds on $\epsilon$
(in the permil range for $1 < m_{Z_d} < 10$~GeV) 
is not possible to saturate the central value of $\Delta a_\mu$ in Eq.~(\ref{eq:amuexp}) in the pure dark-photon 
case ($\epsilon_Z=0$ and $g_d=0$), at least  for $m_{Z_d} > 1$~GeV. 
In the pure dark-$Z$ case ($\epsilon=0$ and $g_d=0$), 
$\Delta a_\mu$ can have the correct magnitude but has the wrong sign. 
As a result, we are forced to have non-vanishing  $U(1)_d$ charges for the muons 
in order to saturate the experimental value of $\Delta a_\mu$
for $m_{Z_d} \sim {\rm few}\times {\rm GeV}$.

\subsubsection*{Effects on Higgs decays}

The possibility to detect $h\to Z Z_d \to 4\ell$ and $h\to Z_d Z_d \to 4\ell$ decays has been extensively discussed in Ref.~\cite{Curtin:2013fra,Davoudiasl:2012ag,Davoudiasl:2013aya}. 
Here we limit ourself to briefly point-out the similarities of the exotic $h\to Z Z_d \to 2\ell 2 \mu$ decay to the SM-allowed 
$h\to Z \Upsilon (\Psi)  \to 2\ell 2 \mu$ processes, and to discuss the possible connection with the $(g-2)_\mu$ anomaly. 

The $h\to Z Z_d$ decay is controlled by the following effective coupling 
\be
\Delta \cL^{(2)}_{\rm int} =  \chZ\, v\, h Z_d^\mu Z_\mu~, 
\ee
generated after electroweak symmetry breaking.
Assuming no $U(1)_d$ charge for the Higgs boson,
the expression of $c_H$ in terms of the mixing parameters $\epsilon$ and $\epsilon_Z$ is 
\ba
\chZ ~ \approx~ 2 \epsilon_Z \frac{m_Z^2}{v^2} + 2 \epsilon \frac{m_{Z_d}^2}{v^2} \tan\theta_W~.
\ea
Comparing this expression with Eq.~(\ref{eq:cLcR})
is quite clear that, contrary to the light-scalar case analyzed in Sect.~\ref{eq:scalar},
the connection between exotic Higgs decays and $\Delta a_\mu$ is more 
model dependent in this framework. The effective coupling $\chZ$ is mainly controlled by $\epsilon_Z$,
while the contribution to $\Delta a_\mu$ is controlled by $g_d Q_{\mu_{L(R)}}^d$.
Still, the $h\to Z Z_d \to 4\ell$ process can provide a very useful constraint on the 
parameter space of the model. In the limit $m_{Z_d}/m_h \ll1$ we have
\ba
\BR (h\to Z Z_d) = \frac{\chZ^2 }{64\pi} ~ \frac{ m_h}{\Gamma_ h}   \frac{v^2  (1-\hrho)^3}{\hrho~ m^2_{Z_d}} ~\approx 
~ 1.9  \times 10^{-4} \left( \frac{\chZ}{10^{-4}} \frac{10~\rm GeV}{m_{Z_d}} \right)^2~.
\ea
For $m_{Z_d} > 1$~GeV and $c_{L,R}$ values relevant to explain the $(g-2)_\mu$ anomaly, the $Z_d$ boson is not long lived
\ba
\Gamma_{Z_d} \ge~ \Gamma(Z_d \to \mu^+\mu^-) = \frac{m_{Z_d}}{24\pi} \left( c_L^2 + c_R^2 \right)
\approx ~ (1.3~{\rm MeV})\times  \frac{m_{Z_d}}{10~\rm GeV} \frac{ c_L^2 + c_R^2 }{0.1^2} ~. 
\ea
As a result, the kinematics of the $h\to Z Z_d \to 4\ell$ decay is identical to that of $h\to Z \Upsilon (\Psi) \to 2\ell 2 \mu$  (with an appropriate shift in the height and the position of the peak in the $m_{34}$ distribution).

The impact on the total $h\to 4\mu$ rate can be written as 
\ba
\frac{ \BR( h \to 4 \mu)_{Z_d} }{ \BR(h \to 4 \mu)_{\rm SM} } & \approx &  0.2 \left( \frac{\chZ}{ 10^{-4}} \frac{10~\rm GeV}{m_{Z_d}} \right)^2 \BR(Z_d \to \mu^+\mu^-)~.
\ea
However, as already discussed in the  light-scalar case, the most efficient way to put bounds on this exotic decay mode is 
by means of the $d\Gamma(h\to 4\mu)/dm_{34}$ distribution. The non-observation of a peak in the present experimental 
 $h\to4\ell$ analyses was used in Ref.~\cite{Curtin:2013fra} to extract the following 95\% C.L.~limit \footnote{~This limit applies to the 
 combination of both muon and electron channels}
\ba
\BR(h\to ZZ_d)\times \BR(Z_d\to \ell\ell) ~ \lesssim  ~ 10^{-4} - 10^{-3}~,
\label{eq:Curt_bound}
\ea
for $12~{\rm GeV} <  m_{Z_d} < 34~{\rm GeV}$, assuming SM Higgs production rate and $\Gamma_{Z_d}\ll 1$ GeV. 
As noted in Ref.~\cite{Curtin:2013fra}, dedicated analysis are needed to search for lighter $Z_d$. 

For illustrative purposes, let's denote the possible future bound on $\BR(h\to ZZ_d)\times \BR(Z_d\to \mu\mu)$ 
as follows
\ba
\BR(h\to ZZ_d)\times \BR(Z_d\to \mu\mu) ~ <  ~ \kappa \times 10^{-5}~.
\label{eq:kappa_bound}
\ea
Using this result 
we would be able to impose the following non-trivial constraint on the effective couplings of the model
\be
0< \left( \frac{\chZ}{ 10^{-4}} \right)^2 \frac{m_{Z_d}}{10~\rm GeV} \frac{\Gamma_h^{\rm SM}}{\Gamma_{Z_d}} \frac{\Delta a_\mu}{2.9 \times 10^{-9}} 
 \frac{c_V^2 + c_A^2 }{c_V^2 - 5 c_A^2}  ~ < ~  0.3 \times \kappa~.
\ee
 Assuming $c_A\ll c_V$ and requiring the $(g-2)_\mu$ anomaly to be explained by this light vector particle, we would find
\be
0< \left( \frac{\chZ}{ 10^{-4}} \right)^2 \frac{m_{Z_d}}{10~\rm GeV} \frac{\Gamma_h^{\rm SM}}{\Gamma_{Z_d}} \lesssim ~ 0.3 \times \kappa~.
\ee
for the range of $m_{Z_d}$ where the bound in Eq.~(\ref{eq:kappa_bound}) holds.

\section{Conclusions}

The $h\to 4\ell$ decays represent a precious source of information about 
the nature of the Higgs boson and, more generally, a sensitive probe of physics 
beyond the SM.  This is particularly true given the  kinematical closure 
of the $ZZ$ threshold. This fact increase the NP sensitivity of the light 
dilepton mass spectrum ($m_{34}$), that can be used to probe 
the existence of non-standard (nearby or distant) poles contributing to 
the $h\to 4\ell$ decay amplitude~\cite{Isidori:2013cla,Grinstein:2013vsa,Buchalla:2013mpa}.

In this paper we have analyzed the possibility to discover 
light poles, within the accessible kinematical range of 
the $d \Gamma(h\to 4\ell )/dm_{34}$ spectrum.
Such spectrum is very sensitive to possible new states singlet 
under the SM gauge group, weakly coupled to Higgs and light leptons.
As we have shown by means of two explicit NP constructions, with new light 
scalar or vector fields, a motivation for the existence of these exotic states is provided by the $(g-2)_\mu$ anomaly.
In a wide region of parameter space relevant to explain the 
$(g-2)_\mu$ anomaly, and consistent with all existing bounds, such states give rise to sizable 
modifications of the $m_{34}$ spectrum in $h\to 4\mu$ (and possibly also 
$h\to 2e2\mu$) decays. These modifications are well within the reach of 
present and future analyses of $d \Gamma(h\to 4\ell )/dm_{34}$ at the LHC.

We have also demonstrated that the $d \Gamma(h\to 4\ell )/dm_{34}$ spectrum is known 
with good theoretical accuracy also in the $m_{34}$ region close to the  
quarkonium thresholds. The latter give rise to tiny effects within the  SM, and 
do not diminish the sensitivity to NP  models with light exotic states in the few GeV range.

\section*{Acknowledgments}
We thank Mario Antonelli and Paride Paradisi for useful discussions.
This work is supported in part by the EU ERC Advanced Grant FLAVOUR (267104), and by MIUR under 
project 2010YJ2NYW.

\end{document}